\documentclass{article}
\usepackage{spconf,amsmath,amssymb,graphicx}
\usepackage{booktabs}
\usepackage{array}
\newcolumntype{C}[1]{>{\centering\arraybackslash}m{#1}}
\usepackage[hidelinks]{hyperref}
\title{ASYMMETRIC DYNAMIC ROUTING: BALANCING REASONING DEPTH AND COMPUTATIONAL EFFICIENCY IN HYPERGRAPH RAG}

\name{Qi Sun\textsuperscript{1*}, Yijia Zhang\textsuperscript{2*}, Caibo Li\textsuperscript{2}, Qiang Li\textsuperscript{3}, Yu Guo\textsuperscript{1\textdagger}\thanks{* Equal contribution. \textdagger\ Corresponding author: yu.guo@xjtu.edu.cn}}
\address{\textsuperscript{1}School of Software Engineering, Xi'an Jiaotong University, Xi'an, Shaanxi, China\\
\textsuperscript{2}State Key Laboratory of Human-Machine Hybrid Augmented Intelligence,\\ and Institute of Artificial Intelligence and Robotics, Xi'an Jiaotong University, Xi'an, Shaanxi, China\\
\textsuperscript{3}EHV Power Transmission Company of China Southern Power Grid Co., Ltd.
}

\begin{document}
\ninept
\setlength{\parindent}{1em}
\maketitle

\begin{abstract}
While graph-based and hypergraph-based Retrieval-Augmented Generation (RAG) significantly mitigate hallucinations in Large Language Models (LLMs), existing structure-based RAG systems typically adopt static traversal strategies regardless of the query complexity. We identify this ``static retrieval fallacy'' as a primary source of computational redundancy for simple queries and cognitive context gaps for complex reasoning tasks. To balance reasoning quality and inference efficiency, we propose Asymmetric Dynamic Routing (ADR), an intent-conditioned retrieval framework operating over hierarchical knowledge graphs. ADR employs a lightweight structured classifier to dynamically dispatch queries among three asymmetric topological traversal operators: localized fact anchoring, bottom-up adjacency diffusion, and top-down insight grounding, which collectively enable bidirectional information flow across hierarchical knowledge layers. Extensive empirical evaluations across five domain-specific corpora demonstrate that ADR maintains strong reasoning performance while reducing prompt token consumption by up to 48.7\% and end-to-end query latency by 45.3\%, yielding a favorable quality--efficiency trade-off for query-adaptive Hypergraph RAG.
\end{abstract}


\section{INTRODUCTION}
\noindent Retrieval-Augmented Generation (RAG) provides non-parametric memory for Large Language
Models (LLMs), grounding generated responses in verifiable external corpora to suppress factual
hallucinations \cite{ref1,ref2,ref3}. To overcome the semantic fragmentation and logical
disconnection inherent in traditional dense vector similarity search, recent approaches have shifted
toward structured knowledge representations. Frameworks such as GraphRAG \cite{ref4}, LightRAG
\cite{ref5}, and Hyper-RAG \cite{ref6} organize domain knowledge into topological graphs and
hypergraphs. By modeling explicit entity dependencies and higher-order semantic associations, these
structured representations provide LLMs with multi-hop deductive pathways, enabling comprehensive
synthesis across disparate documents \cite{ref7,ref8}.

However, despite these topological advancements, existing stucture-based RAG systems still rely on uniform
configurations for subgraph expansion radius and traversal depth~\cite{ref9}, regardless of query
complexity. We define this limitation as the \textit{Static Retrieval Fallacy}, which leads to two
contrasting problems:

\begin{itemize}
    \item \textbf{Context Pollution via Over-Retrieval:} For simple factual queries, unnecessary
    multi-hop diffusion introduces topologically adjacent but semantically irrelevant information
    into the context. This degrades the signal-to-noise ratio within the LLM's cross-attention
    layers, increases prefill memory consumption, and may induce associative
    hallucinations~\cite{ref10}.

    \item \textbf{Cognitive Gaps via Under-Retrieval:} Abstract, cross-document inquiries require
    macro-level synthesis. Localized topological expansion may fail to bridge disparate abstraction
    layers, resulting in fragmented evidence and superficial reasoning rather than mechanistic
    explanation~\cite{ref11}.
\end{itemize}

Iterative, reflection-driven methods such as Self-RAG~\cite{ref12}, IRCoT~\cite{ref13}, and
CogRAG~\cite{ref14} introduce adaptability through multi-turn agentic loops. However, they incur
multi-second inference latency overheads, limiting their applicability to high-throughput, real-time
deployment in speech and signal processing systems.

To resolve the tension between reasoning depth and computational efficiency, we propose
\textbf{Asymmetric Dynamic Routing (ADR)}, a single-pass, intent-guided framework that dynamically
adjusts the retrieval entry level, traversal direction, and expansion scope over a hierarchical
knowledge substrate.

The main contributions of this work are threefold:
\begin{itemize}
    \item We formalize the \textit{Static Retrieval Fallacy}, highlighting the mismatch between
    uniform graph traversal and heterogeneous query complexity.

    \item We develop ADR as a query-conditioned framework for bidirectional topological traversal, 
    comprising three asymmetric operators: localized fact anchoring, bottom-up adjacency diffusion, 
    and top-down insight grounding.

    \item Through evaluations across five distinct domains, we find that traversal requirements vary across 
    domains and demonstrate that ADR maintains strong reasoning performance while substantially reducing 
    prompt token consumption and latency.
\end{itemize}

\section{Related Work}

\noindent \textbf{Stucture-based RAG Architectures.}
The transition from flat chunk retrieval to topological indexing represents a major shift in RAG
architectures. GraphRAG~\cite{ref4} uses LLM-extracted knowledge graphs to generate hierarchical
community summaries, LightRAG~\cite{ref5} adopts dual-level retrieval, and GNN-RAG~\cite{ref15}
applies graph neural networks to retrieve relevant reasoning paths. Hyper-RAG~\cite{ref6} further
uses hypergraph representations to capture pairwise and beyond-pairwise correlations. Despite these
advances, their retrieval procedures remain largely predefined rather than conditioned on query
complexity, potentially causing unnecessary expansion for simple queries and insufficient relational
coverage for complex ones.

\textbf{Adaptive and Agentic Retrieval Paradigms.}
Recent studies explore query-adaptive mechanisms to address static retrieval inefficiencies.
Adaptive-RAG~\cite{ref16} selects among no-retrieval, single-step, and iterative retrieval according
to query complexity; Self-RAG~\cite{ref12} uses learned reflection tokens to control retrieval;
ARAG~\cite{ref17} triggers retrieval based on low-confidence predictions; and CRAG~\cite{ref18} evaluates 
retrieval quality to initiate corrective actions. While powerful, these agentic and iterative loops may 
propagate generation errors and introduce significant temporal latency. Moreover, existing adaptive methods 
primarily vary retrieval frequency or reasoning depth, without explicitly adapting the topological traversal 
to heterogeneous query requirements.

\section{PROPOSED METHODOLOGY}

\begin{figure*}[t]
\centering
\includegraphics[width=\textwidth]{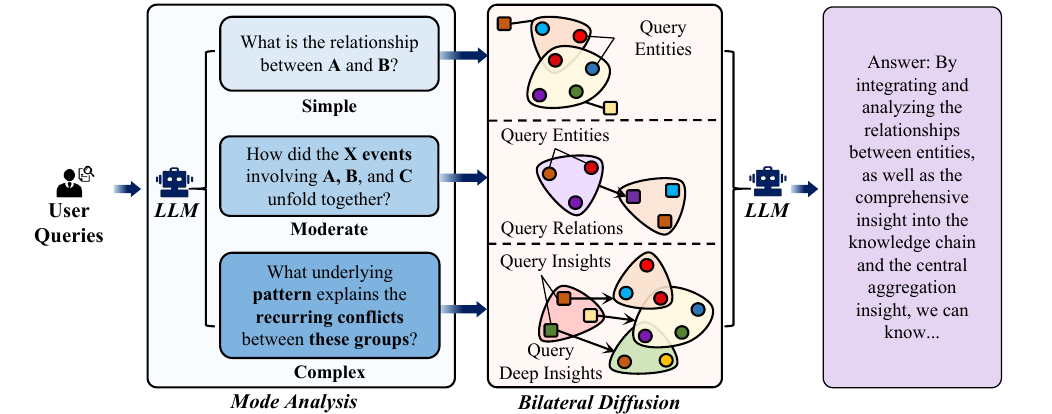}
\caption{The architectural blueprint of Asymmetric Dynamic Routing (ADR).}
\label{fig:architecture}
\end{figure*}

\noindent As shown in Fig.~\ref{fig:architecture}, incoming queries are classified into three
categories based on their retrieval and reasoning requirements:
\begin{itemize}
\item \textbf{Simple queries:} Trigger localized anchoring within the factual hypergraph $H_K$.
\item \textbf{Moderate queries:} Execute bottom-up diffusion into the deep insight hypergraph $H_D$.
\item \textbf{Complex queries:} Initiate top-down semantic matching over $H_D$, followed by a
inverse cross-layer mapping $\pi^{-1}$ to ground the high-level insights in verifiable supporting
factual evidence within $H_K$.
\end{itemize}

\subsection{Hierarchical Dual-Hypergraph Representation}

\noindent ADR operates on the dual-hypergraph knowledge framework introduced in DHI~\cite{ref19} and
follows its original construction procedure. Given a document corpus, the knowledge space is
organized as $\mathcal{G}=(H_K,H_D)$, where $H_K=(V_K,E_K)$ preserves fine-grained entities and
factual relations directly grounded in the source text, while $H_D=(V_D,E_D)$ organizes higher-level
abstract structures and their semantic associations.

In $H_K$, $V_K$ represents factual entities, while $E_K$ contains pairwise factual relations
($|e|=2$) and higher-order factual hyperedges ($|e|\geq3$). The latter capture joint relations among
multiple entities and provide the structural basis for higher-level abstraction. Each higher-order
factual hyperedge is associated with an atomic direct insight $\iota(e)$ and mapped by $\pi$ to a
corresponding insight vertex in $V_D$, whereas pairwise edges remain within $H_K$. Each deep insight
edge $\varepsilon\in E_D$ connects a set of insight vertices and carries a synthesized narrative
insight $I(\varepsilon)$. ADR uses this cross-layer correspondence to adapt the retrieval level and
traversal direction to query complexity.

\subsection{Intent Routing and Conservative Fallback}

\noindent ADR categorizes an incoming query $q$ into three query categories according to its
retrieval and reasoning requirements, with $\mathcal{Y}=\{1,2,3\}$:

\begin{itemize}
    \item \textbf{Simple ($y=1$):} Targets distinct entities, definitions, or direct factual
    relations that can be resolved through localized factual retrieval.

    \item \textbf{Moderate ($y=2$):} Requires relational or procedural reasoning across multiple
    facts, with localized evidence supplemented by broader contextual associations.

    \item \textbf{Complex ($y=3$):} Requires high-level thematic synthesis, cross-domain comparison,
    or causal reasoning based on abstract semantic structures and supporting factual evidence.
\end{itemize}

ADR employs an LLM-based intent classifier $f_{\phi}$ to output a predicted category
$y\in\mathcal{Y}$ and confidence $c_y\in[0,1]$. To handle low-confidence predictions, ADR adopts a
Conservative Fallback Principle: the prediction is accepted when $c_y\geq\gamma$; otherwise, the
query is redirected to Moderate, as defined in Eq.~\eqref{eq:routing}:
\begin{equation}
\label{eq:routing}
\hat{y}
=
y\,\mathbb{I}(c_y\geq\gamma)
+
2\,\mathbb{I}(c_y<\gamma),
\end{equation}
where $\hat{y}$ denotes the final routing decision and $\gamma=0.75$ by default. Moderate is
selected as the fallback because it provides an intermediate retrieval scope, reducing the risk of
routing uncertain queries to either extreme while maintaining bounded retrieval cost.

\subsection{Asymmetric Bidirectional Traversal Operators}

\subsubsection{Operator 1: Localized Fact Anchoring ($\hat{y}=1$)}

\noindent For Simple queries, ADR performs localized retrieval within the factual hypergraph $H_K$,
as the required evidence can typically be obtained directly from relevant entities and factual
relations. Given a query $q$, ADR identifies the $k_e$ most relevant factual entities to form the
entity anchor set $V_{\mathrm{eas}}$ in Eq.~\eqref{eq:anchors}:
\begin{equation}
\label{eq:anchors}
V_{\mathrm{eas}}
=
\underset{
V'\subset V_K,\ |V'|=k_e
}{\arg\max}
\sum_{v\in V'}
\cos(\mathbf{e}_v,\mathbf{q}),
\end{equation}
where $\mathbf{e}_v$ denotes the embedding of factual entity $v$. ADR then retrieves the factual
edges directly associated with these anchors using Eq.~\eqref{eq:factual-edges}:
\begin{equation}
\label{eq:factual-edges}
E_{\mathrm{sub}}
=
\{e\in E_K \mid e\cap V_{\mathrm{eas}}\neq\emptyset\}.
\end{equation}

Retrieval terminates at $E_{\mathrm{sub}}$ without further neighborhood expansion or cross-layer
traversal. The granular factual content associated with $E_{\mathrm{sub}}$ is then formatted as the
retrieval context for final generation. This restricted expansion provides an effect analogous to
zero-hop retrieval while avoiding unnecessary context expansion.

\subsubsection{Operator 2: Bottom-Up Adjacency Diffusion ($\hat{y}=2$)}

\noindent For Moderate queries, directly anchored factual evidence may be insufficient to capture
the broader relational context required for reasoning. ADR therefore extends retrieval upward from
$H_K$ to $H_D$. Since $\pi$ is defined only for higher-order factual hyperedges, those in
$E_{\mathrm{sub}}$ are mapped to their corresponding insight vertices using
Eq.~\eqref{eq:insight-seeds}:
\begin{equation}
\label{eq:insight-seeds}
U_{\mathrm{seed}}
=
\{\pi(e)\mid e\in E_{\mathrm{sub}},\ |e|\geq3\}.
\end{equation}

Starting from $U_{\mathrm{seed}}$, ADR performs a single adjacency expansion in $H_D$ to retrieve
the associated deep insight edges in Eq.~\eqref{eq:upward-edges}:
\begin{equation}
\label{eq:upward-edges}
E_{\mathrm{up}}
=
\{\varepsilon\in E_D
\mid
\varepsilon\cap U_{\mathrm{seed}}\neq\emptyset\}.
\end{equation}

The granular factual content associated with $E_{\mathrm{sub}}$ and the synthesized narrative
insights associated with $E_{\mathrm{up}}$ are then jointly formatted as the retrieval context. This
bottom-up traversal preserves factual grounding while extending the context to broader semantic
associations required by Moderate queries.

\subsubsection{Operator 3: Top-Down Insight Grounding ($\hat{y}=3$)}

\noindent For Complex queries, localized factual anchoring may fail to capture the high-level
semantic structures required for thematic synthesis, comparison, or causal reasoning. ADR therefore
reverses the traversal direction and starts from $H_D$, retrieving the $k_d$ deep insight edges most
relevant to the query using Eq.~\eqref{eq:insight-retrieval}:
\begin{equation}
\label{eq:insight-retrieval}
E^{*}
=
\underset{
E'\subset E_D,\ |E'|=k_d
}{\arg\max}
\sum_{\varepsilon\in E'}
\cos(\mathbf{e}_{\varepsilon},\mathbf{q}),
\end{equation}
where $\mathbf{e}_{\varepsilon}$ denotes the embedding of deep insight edge $\varepsilon$. The
retrieved insight structures are then grounded in their supporting factual evidence through the
inverse cross-layer mapping in Eq.~\eqref{eq:grounding}:
\begin{equation}
\label{eq:grounding}
E_{\mathrm{ground}}
=
\bigcup_{\varepsilon\in E^{*}}
\{\pi^{-1}(u)\mid u\in\varepsilon\}
\subseteq E_K.
\end{equation}

The synthesized narrative insights associated with $E^{*}$ and the granular factual content
associated with $E_{\mathrm{ground}}$ are jointly formatted as the retrieval context. This top-down
traversal combines high-level semantic reasoning with supporting factual evidence from $H_K$.

\section{EXPERIMENTAL EVALUATION}

\subsection{Experimental Configuration}

\begin{table}[tb]
\centering
\fontsize{9}{11.5}\selectfont
\setlength{\tabcolsep}{1pt}
\renewcommand{\arraystretch}{1.05}
\caption{Statistical information of the datasets.}
\label{tab:1}
\begin{tabular}{@{}C{.19\columnwidth}*{5}{C{\dimexpr(.81\columnwidth-10pt)/5\relax}}@{}}
\toprule
Statistics & Mix & CS & Agri. & Neuro. & Patho. \\
\midrule
Documents & 61 & 10 & 12 & 1 & 1 \\
Chunks & 560 & 1,992 & 1,813 & 1,790 & 824 \\
Tokens & 615,355 & 2,190,803 & 1,993,515 & 1,968,716 & 905,760 \\
\bottomrule
\end{tabular}
\end{table}

\noindent \textbf{Datasets.}
We evaluate ADR on five datasets: Mix, CS, and Agriculture from UltraDomain~\cite{ref20}, and
Neurology and Pathology from MIRAGE~\cite{ref21}. Their statistics are summarized in
Table~\ref{tab:1}. Mix represents a cross-domain sparse setting; CS and Agriculture represent
intra-domain sparse settings with relatively weak inter-passage dependencies; and Neurology and
Pathology represent intra-domain dense settings with stronger semantic continuity and local
relations.

\textbf{Baselines and Protocols.} We compare ADR with six representative baselines covering
non-retrieval, conventional RAG, graph-based RAG, and hypergraph-based RAG methods:

\begin{itemize}
\item \textbf{LLM (Zero-shot):} Directly generates responses without external knowledge retrieval.
\item \textbf{NaiveRAG \cite{ref1}:} Retrieves relevant text chunks through dense vector similarity
search.
\item \textbf{GraphRAG \cite{ref4}:} Organizes knowledge into entity-relation graphs and supports
global retrieval through hierarchical community summaries.
\item \textbf{LightRAG \cite{ref5}:} Employs a dual-level retrieval framework that integrates
entity-level and relation-level information.
\item \textbf{Hyper-RAG \cite{ref6}:} Uses hypergraph structures to model higher-order relations
among multiple entities.
\item \textbf{DHI \cite{ref19}:} Constructs coupled factual and deep insight hypergraphs and
performs cross-layer retrieval to integrate fine-grained facts with higher-level insights.
\end{itemize}
All methods use GPT-4o-mini as the generation backbone and \texttt{text-embedding-3-small} for
representation, with temperature $T=0$ for deterministic evaluation. We adopt a normalized
LLM-as-a-Judge protocol (0--100) across five dimensions: Comprehensiveness, Diversity, Empowerment,
Logicality, and Readability.

\subsection{Overall Performance Comparison}

\begin{table}[tb]
\centering
\fontsize{9}{11.5}\selectfont
\setlength{\tabcolsep}{1pt}
\renewcommand{\arraystretch}{1.05}
\caption{Overall performance comparison with baseline methods across five datasets (0--100 scale,
higher is better).}
\label{tab:2}
\begin{tabular}{@{}C{.32\columnwidth}*{5}{C{\dimexpr(.68\columnwidth-10pt)/5\relax}}@{}}
\toprule
Method & Mix & CS & Agri. & Neuro. & Patho. \\
\midrule
LLM (Zero-shot) & 79.30 & 81.08 & 79.64 & 81.15 & 82.81 \\
NaiveRAG \cite{ref1} & 78.09 & 79.43 & 76.20 & 79.20 & 82.04 \\
GraphRAG \cite{ref4} & 81.06 & 84.03 & 79.98 & 83.10 & 82.72 \\
LightRAG \cite{ref5} & 81.01 & 81.25 & 79.05 & 81.82 & 84.43 \\
Hyper-RAG \cite{ref6} & 80.39 & 83.88 & 81.98 & 83.74 & 84.41 \\
DHI \cite{ref19} & 83.18 & 84.66 & 82.97 & 84.23 & 85.78 \\
\textbf{ADR (Ours)} & \textbf{83.91} & \textbf{85.24} & \textbf{83.86} & \textbf{85.02} & \textbf{86.65} \\
\bottomrule
\end{tabular}
\end{table}

\noindent Table~\ref{tab:2} compares ADR with six baseline methods across the five datasets. ADR
achieves consistently strong performance across different knowledge settings. In particular, ADR
obtains scores of 83.91 on Mix and 86.65 on Pathology, compared with 83.18 and 85.78 for DHI,
respectively. Similar improvements are observed on CS, Agriculture, and Neurology, indicating that
query-adaptive traversal can support heterogeneous retrieval requirements while maintaining stable
performance across domains.

\subsection{Dynamic Routing Analysis and Computational Efficiency}

\noindent To examine the effectiveness of dynamic routing, we compare ADR with three fixed traversal
modes: Simple-only, Moderate-only, and Complex-only. Each fixed configuration applies the same
retrieval strategy to all queries, whereas ADR selects the retrieval path according to query
complexity.

As shown in Table~\ref{tab:3}, ADR outperforms the best fixed configuration by 1.39, 2.11, 1.16, 0.94, and 1.25 points on Mix, CS, Agriculture, Neurology, and Pathology, respectively. The relative
effectiveness of fixed traversal modes also varies across datasets with different knowledge
characteristics. On CS, Complex-only achieves 83.13 compared with 82.14 for Moderate-only,
suggesting that queries in this sparse setting may benefit from deeper retrieval. In contrast,
Moderate-only outperforms Complex-only by 1.39 and 1.51 points on Agriculture and Pathology,
respectively, indicating that additional traversal depth does not always improve response quality.
These results show that no fixed traversal mode is consistently optimal across different knowledge
settings. This adaptive mechanism enables ADR to preserve localized evidence for simpler queries
while invoking deeper cross-layer retrieval when additional reasoning context is required.

ADR also improves computational efficiency. As shown in Table~\ref{tab:4}, ADR reduces average
latency from 435 to 238 ms and prompt tokens from 3,760 to 1,929 relative to Complex-only,
corresponding to reductions of 45.3\% and 48.7\%, respectively, while improving the overall score
from 82.05 to 83.91.

\begin{table}[tb]
\centering
\fontsize{9}{11.5}\selectfont
\setlength{\tabcolsep}{1pt}
\renewcommand{\arraystretch}{1.05}
\caption{Ablation study of different routing strategies across five datasets.}
\label{tab:3}
\begin{tabular}{@{}C{.35\columnwidth}*{5}{C{\dimexpr(.65\columnwidth-10pt)/5\relax}}@{}}
\toprule
Routing Category & Mix & CS & Agri. & Neuro. & Patho. \\
\midrule
Frozen: Simple-only & 82.01 & 82.09 & 81.89 & 83.41 & 83.40 \\
Frozen: Moderate-only & 82.52 & 82.14 & 82.70 & 84.08 & 85.40 \\
Frozen: Complex-only & 82.05 & 83.13 & 81.31 & 83.76 & 83.89 \\
\textbf{ADR (Dynamic)} & \textbf{83.91} & \textbf{85.24} & \textbf{83.86} & \textbf{85.02} & \textbf{86.65} \\
$\Delta$ gain over best static & \textbf{+1.39} & \textbf{+2.11} & \textbf{+1.16} & \textbf{+0.94} & \textbf{+1.25} \\
\bottomrule
\end{tabular}
\end{table}

\begin{table}[tb]
\centering
\fontsize{9}{11.5}\selectfont
\setlength{\tabcolsep}{1pt}
\renewcommand{\arraystretch}{1.05}
\caption{Computational efficiency profiling and resource parsimony.}
\label{tab:4}
\begin{tabular}{@{}C{.40\columnwidth}*{3}{C{\dimexpr(.60\columnwidth-6pt)/3\relax}}@{}}
\toprule
Routing Category & Latency & Tokens & Score \\
\midrule
Fixed Complex-only & 435~ms & 3,760 & 82.05 \\
\textbf{ADR (Dynamic)} & \textbf{238~ms} & \textbf{1,929} & \textbf{83.91} \\
$\Delta$ Savings & \textbf{-45.3\%} & \textbf{-48.7\%} & \textbf{+1.86} \\
\bottomrule
\end{tabular}
\end{table}

\subsection{Sensitivity Analysis of the Routing Threshold}

\noindent When $\gamma$ is too low, ambiguous queries with relatively low confidence may still be
accepted as reliable routing decisions rather than triggering the conservative fallback.
Consequently, they may be assigned to inappropriate retrieval trajectories with either insufficient
or excessive retrieval depth, resulting in missing evidence or unnecessary context. For example, on
CS, reducing $\gamma$ from 0.75 to 0.50 decreases the Logicality score by 1.14 points. Conversely,
an overly high threshold ($\gamma \geq 0.85$) causes 68.4\% of queries to fall back to the Moderate
trajectory, increasing average latency by 58 ms without significant performance gains ($p>0.05$).
Overall, $\gamma=0.75$ provides a favorable balance between routing reliability and computational
efficiency and is used as the default setting.
\subsection{Qualitative Analysis of Reasoning Pathways}

\noindent We further examine a complex causal query from the Pathology corpus asking how somatic IDH1 mutations promote gliomagenesis through epigenetic remodeling. Simple-only retrieves localized facts about IDH1 and the R132H mutation but fails to capture the further causal chain, whereas Complex-only retrieves 3,850 tokens of tangential biochemical context, including irrelevant associations such as EGFR amplification.

ADR classifies the query as Complex ($c_y=0.94$). Starting from the deep insight hypergraph $H_D$, Top-Down Insight Grounding identifies the causal mechanism: mutant IDH1 drives 2-hydroxyglutarate (2-HG) accumulation, which inhibits $\alpha$-KG-dependent dioxygenases and induces further epigenetic remodeling. ADR then applies $\pi^{-1}$ to ground this mechanism in supporting facts from $H_K$, including the IDH1 R132H mutation and its altered metabolic activity. This cross-layer retrieval preserves the causal progression from mutation to metabolic disruption and epigenetic remodeling while filtering unrelated associations, yielding a more causally coherent response with fewer hallucinated associations and 41.2\% fewer prompt tokens than Complex-only.

\section{Discussion and Limitations}

\noindent While ADR balances retrieval efficiency and reasoning fidelity, several limitations
remain. First, Intent Routing operates independently of the real-time graph structure. For a Complex
query in a structurally sparse domain, insufficient graph connectivity may limit the supporting
factual evidence retrieved by Top-Down Insight Grounding, requiring a fallback to dense vector
search. Second, constructing the dual-hypergraph framework $(H_K,H_D)$ incurs upfront indexing
costs, reflecting a trade-off between offline construction and online inference efficiency.

\section{Conclusion}

\noindent In this paper, we introduced Asymmetric Dynamic Routing (ADR) to address the ``static
retrieval fallacy'' in stucture-based RAG Architectures. ADR adapts retrieval depth and traversal direction to
query complexity through three asymmetric bidirectional traversal operators: localized fact
anchoring, bottom-up adjacency diffusion, and top-down insight grounding. Experiments across five
domains demonstrate that query-adaptive routing maintains strong reasoning performance while
reducing unnecessary traversal and context overhead, with latency and prompt token consumption
reduced by 45.3\% and 48.7\%, respectively. Future work will explore continuous soft routing over
dynamic multi-modal hypergraphs.

\vfill\pagebreak

\bibliographystyle{IEEEtran}
\bibliography{ADRref}
\end{document}